# Growth and electrical characterization of $Al_{0.24}Ga_{0.76}As/Al_xGa_{1-x}As/Al_{0.24}Ga_{0.76}As$ modulation-doped quantum wells with extremely low x

Geoffrey C. Gardner[1,2], John D. Watson[1,3], Sumit Mondal[1,3], Nianpei Deng[3], Gabor A. Csáthy[3] and Michael J. Manfra[1,2,3,4]

[1] Birck Nanotechnology Center, Purdue University, West Lafayette IN, 47907

[2] School of Materials Engineering, Purdue University, West Lafayette IN, 47907

[3] Department of Physics, Purdue University, West Lafayette IN, 47907

[4] School of Electrical and Computer Engineering, Purdue University, West Lafayette IN, 47907

**We report on the growth and electrical characterization of modulation-doped $Al_{0.24}Ga_{0.76}As/Al_xGa_{1-x}As/Al_{0.24}Ga_{0.76}As$ quantum wells with mole fractions as low as x=0.00057. Such structures will permit detailed studies of the impact of alloy disorder in the fractional quantum Hall regime. At zero magnetic field, we extract an alloy scattering rate of 24 ns$^{-1}$ per %Al. Additionally we find that for x as low as 0.00057 in the quantum well, alloy scattering becomes the dominant mobility-limiting scattering mechanism in ultra-high purity two-dimensional electron gases typically used to study the fragile $\nu=5/2$ and $\nu=12/5$ fractional quantum Hall states**.

Presently the fractional quantum Hall effect (FQHE) in the 2$^{nd}$ Landau level is under intense scrutiny [1-14]. It is speculated that the exotic fractional states at filling factors $\nu=5/2$ and $\nu=12/5$ may support low-lying charged excitations that obey non-Abelian braiding statistics [15-19]. For particles obeying non-Abelian statistics repeated interchange of two identical particles does not change the many-body wavefunction by a factor of +/-1 as for bosons and fermions respectively, but rather, results in a unitary transformation of the wavefunction within a degenerate manifold. If the $\nu=5/2$ and $\nu=12/5$ states do indeed support non-Abelian excitations, they may provide a viable platform for quantum computation that is topologically protected from decoherence. However, excitation gap energies for FQHE states in the 2$^{nd}$ Landau level (LL) are typically quite small, presumably limited by disorder. The largest gap measured at $\nu=5/2$ amounts to $\Delta=570$mK and the gap at 12/5 is below 100mK [3, 4, 20, 36] while the theoretical estimate for the 5/2 gap in the density range of current experiments is $\Delta=1.8$K [21-23]. It is, therefore, of considerable interest to understand how different types of disorder (e. g. long-range Coulomb scattering, short-range alloy disorder, and interface roughness scattering) impact the measured excitation gaps [7, 24-26].

Alloy disorder scattering occurs when electrons traverse a region of semiconductor comprised of a random solution of two or more binary semiconductors. $Al_xGa_{1-x}As$ is such a random alloy in which x is the mole fraction of aluminum in solution.  In the case of $Al_xGa_{1-x}As$, alloy disorder scattering is essentially short-ranged, arising from the replacement of a Ga atom with an isovalent Al atom and is described in Ref. [27].  It is operative on the scale of the unit cell and can be represented as a sum of delta-function scattering potentials.  Li et al. [28] studied alloy scattering of two-dimensional electrons in $Al_{0.33}Ga_{0.67}As/Al_xGa_{1-x}As$ single heterojunctions.  They determined an alloy scattering rate of $35ns^{-1}$ per % Al and an alloy scattering potential $U$=1.13eV.  Importantly, Li and collaborators used these samples to establish scaling and universality of the *integer* quantum Hall plateau-to-plateau transitions [29].  However, the heterostructure design described in Ref. [28] does not produce samples of sufficient quality to study the fragile FQHE of the 2$^{nd}$ LL.  Additionally, samples with even lower alloy content in a quantum well and over a broader range of x are necessary for studies of the 2$^{nd}$ LL.

In this letter we describe the growth and electrical characterization of modulation-doped $Al_{0.24}Ga_{0.76}As/Al_xGa_{1-x}As/Al_{0.24}Ga_{0.76}As$ quantum wells with alloy levels down to x=0.00057 incorporated into a modern heterostructure design that is typically employed in studies of the FQHE in the 2$^{nd}$ LL. Samples were grown by molecular beam epitaxy (MBE) in a customized system specifically designed to grow ultra-high purity structures necessary to study the FQHE in the 2$^{nd}$ Landau level. This system has now produced many samples with electron mobility exceeding $20x10^6 cm^2/Vs$ and activation gaps for the ν=5/2 state Δ>500mK.  The MBE is configured with 2 aluminum and 2 gallium effusion cells so that heterostructures containing multiple values of alloy mole fraction x can be grown without changing the effusion cell temperatures during growth.  Reflection high energy electron diffraction (RHEED) intensity oscillations collected with a computer controlled CCD camera were analyzed to calibrate growth rates and the aluminum content in the barriers and quantum wells. A series of 10 samples with varying x in the quantum well (including x=0) were grown.

Our alloy-disorder samples are modulation doped $Al_{0.24}Ga_{0.76}As/Al_xGa_{1-x}As/Al_{0.24}Ga_{0.76}As$ quantum well structures, consisting of a 30nm $Al_xGa_{1-x}As$ quantum well sandwiched between $Al_{0.24}Ga_{0.76}As$ barriers. The structure is doped with Si at a setback of 75nm above and below the quantum well using a short-period superlattice doping scheme [30-31].  We have found that this doping method consistently yields the largest energy gaps in the 2$^{nd}$ LL, and such large energy gaps are a necessary starting condition before intentionally adding disorder of any kind.  In the 10 samples studied, the Al alloy content in the quantum well was then varied from x=0.0 to x=0.0078.

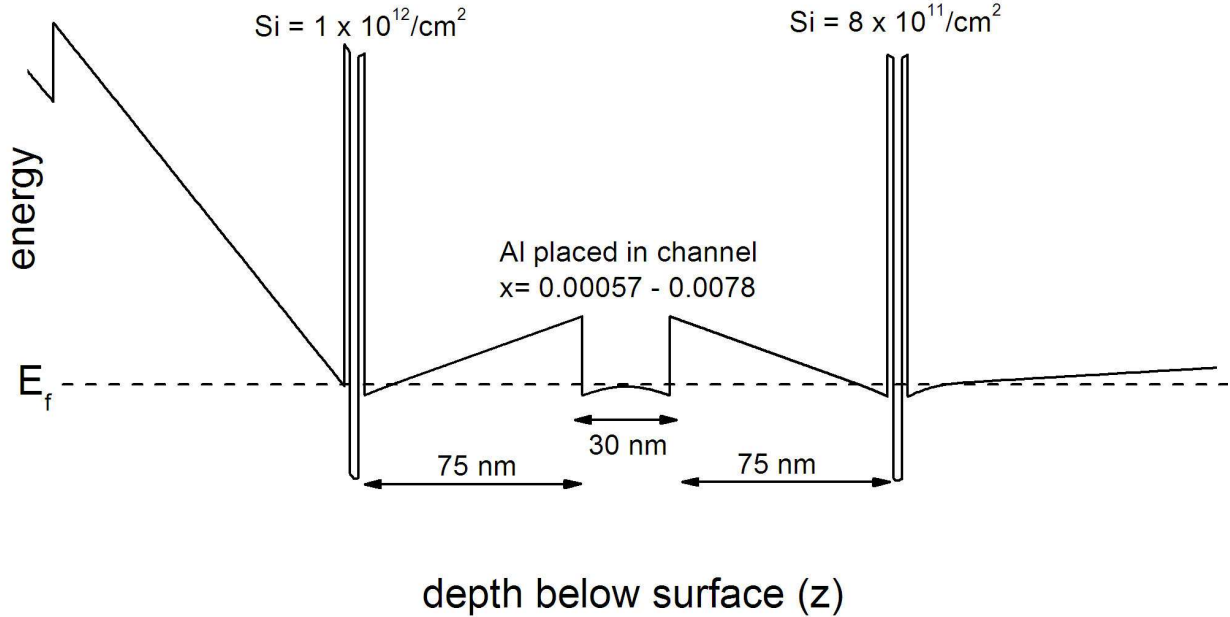

Figure 1: A schematic band structure of the samples used in this study. Note that the Si dopants are placed in narrow GaAs quantum wells 75nm above and below the principal 30nm quantum well [37].

Heterostructures were grown on semi-insulating (100) GaAs substrates using a single Ga and two Al effusion cells. The primary Al cell was used to grow the $Al_{0.24}Ga_{0.76}As$ layers while the secondary Al cell was used for introducing the small concentration of Al in the quantum well. The effusion cells were allowed to stabilize at their approximate growth temperatures for at least 30 minutes prior to fine tuning their respective growth rates via RHEED intensity oscillations. RHEED oscillations were observed along the $[1\bar{1}1]$ direction on the 2x4 reconstructed surface of GaAs with a CCD camera and custom data acquisition software [32]. The long time scale associated with the AlAs oscillations of the secondary Al cell required special precautions. To minimize intensity fluctuations due to background lighting the RHEED screen and CCD camera were enclosed in a light tight box. The angle of the incidence for the RHEED beam was carefully tuned to provide the strongest oscillations. An azimuthal positioning process that controlled the approach direction minimized noise associated with azimuthal rotation gear tolerances of the sample mounting stage. We also found it necessary to reduce the RHEED electron gun filament intensity below its standard operating level to minimize drift of the specular spot due to charging of the substrate. This step allows for data collection from a very small area on the phosphor screen, eliminating background signal from the higher order peaks in the diffraction pattern. These combined techniques allow us to characterize very slow AlAs monolayer formation. A plot of characteristic AlAs RHEED oscillations for x=0.0036 is shown in Figure 2. The period of the oscillations was determined by visual identification of wave minima and maxima. This process was repeated several times before each growth to ensure stability and reproducibility of the aluminum mole fraction in the quantum well. We also note that during the growth of the Al$_{0.24}$Ga$_{0.76}$As barrier grown immediately prior to the quantum well, the secondary Al cell (used for alloy disorder) shutter was opened. By opening the shutter during the barrier the impact of transients in the Al flux on the quantum well caused by the shutter changing state was minimized. This procedure most accurately recreated the RHEED calibration conditions, ensuring that the AlAs growth rate during the deposition of the quantum well matched the rate measured with RHEED oscillations. The additional aluminum deposited in the barrier was negligible.

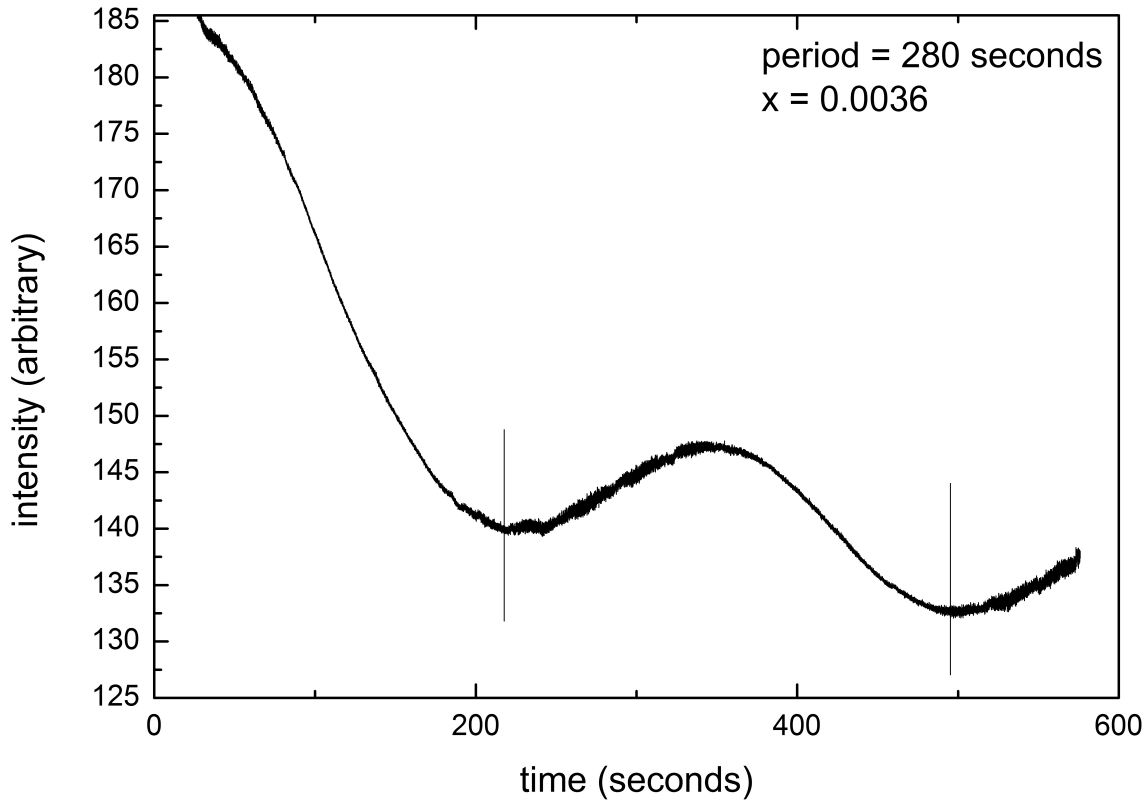

Figure 2: Plot of RHEED oscillations for Al concentration of x=0.0036. The large drop in intensity seen at the start of the data collection is characteristic of AlAs oscillations in our machine. The vertical lines represent deposition of one complete monolayer of AlAs.

As the Al mole fraction was increased from x=0.0 to x=0.0078, the mobility decreased from $16 \times 10^6$ cm$^2$/Vs to $1.2 \times 10^6$ cm$^2$/Vs. Carrier mobility was measured using ~4 mm x 4 mm square samples contacted with In-Sn alloy annealed into the sample at 430°C for 15 min in H$_2$/N$_2$ forming gas. The material was characterized at T=300 mK after illumination with a red light emitting diode using standard lock-in techniques with the density being determined from quantum Hall effect (QHE) minima. The carrier areal density of the samples used in this study shows no dependence on the Al mole fraction in the channel. The aluminum atoms added to the channel have the same number of valence electrons as the gallium atoms they replace and do not act as donors or acceptors.

Table 1 summarizes the results from electrical measurements conducted at T=0.3K. The data clearly demonstrates that mobility, defined as $\mu=e\tau/m^*$ ($\tau$ is the mobility lifetime and $m^*=0.067m_0$ is the effective mass in GaAs), is strongly impacted by the introduction of aluminum in the quantum well. The inverse scattering time due to alloy disorder ($1/\tau_{alloy}$) is expected to be proportional to x(1-x) [27]. In Figure 3 we plot the experimentally-measured inverse total scattering time ($1/\tau_{total}$) vs. x(1-x) along with a linear fit to the data. The total scattering rate (at T=0) is given by Matthiessen's rule as $1/\tau_{total}=1/\tau_{alloy} + 1/\tau_{other}$, where $1/\tau_{other}$ represents scattering from all other temperature independent mechanisms. We neglect phonon scattering as it is well known that phonon scattering does not contribute significantly to the mobility lifetime at T=0.3K [33-35]. An alloy scattering rate of 24ns$^{-1}$ per % Al was determined from the linear fit. We note that our extracted alloy scattering rate differs from the result quoted in Ref. [28] by more than 30%. We attribute this discrepancy to a neglect of a zero offset in Ref. [28]. When we fit the data of Ref. [28] using the methods described here we find a scattering of 26ns$^{-1}$ per %Al, more consistent with our measurements.

TABLE I: Sample identifier, quantum well mole fraction x, electron density $n$, mobility $\mu$, and total scattering rate $\tau^{-1}$ for the 10 samples grown and measured at T=0.3K

| Sample | x | $n$ ($10^{11}$/cm$^2$) | $\mu$ ($10^6$ cm$^2$/V s) | $\tau^{-1}$ (ns$^{-1}$) |
|---|---|---|---|---|
| 1 | 0.0 | 2.92 | 16 | 1.6 |
| 2 | 0.00057 | 2.98 | 6.5 | 4.0 |
| 3 | 0.00075 | 2.90 | 5.0 | 5.2 |
| 4 | 0.00082 | 2.98 | 4.1 | 6.4 |
| 5 | 0.00130 | 2.97 | 3.9 | 6.7 |
| 6 | 0.00150 | 3.00 | 3.6 | 7.3 |
| 7 | 0.00260 | 2.78 | 2.7 | 9.7 |
| 8 | 0.00360 | 3.13 | 2.2 | 12 |
| 9 | 0.00460 | 2.82 | 1.7 | 15 |
| 10 | 0.00780 | 2.80 | 1.2 | 22 |

At this juncture two points merit discussion. The x=0.0 intercept of the linear fit is significant. The linear fit suggests that the x=0.0 intercept is approximately $3ns^{-1}$, while the actual measured value for the sample grown at x=0.0 is $1.6ns^{-1}$. This offset, designated as δ in Fig. 3, is attributed to additional impurities introduced from the 2$^{nd}$ Al effusion cell and the surrounding material when the 2$^{nd}$ Al shutter is opened. This observation indicates that the effusion cells themselves are still sources of impurities and points to a direction for future improvement in material quality. Note that this effect is only observable when starting with extremely high mobility samples. The second point concerns the small amount of alloy disorder necessary for alloy scattering to become the dominant scattering mechanism controlling the mobility lifetime. Our MBE system now routinely produces samples with mobility in excess of $20 \times 10^6 cm^2/Vs$. At $\mu = 20 \times 10^6 cm^2/Vs$ the scattering rate $1/\tau = 1.3 ns^{-1}$. From the data of Fig. 3, it is clear that alloy disorder greater than or equal to x=0.0005 will become the dominant mobility-limiting mechanism in state-of-the-art samples. While alloy disorder clearly impacts mobility, its influence on the FQHE in the 2$^{nd}$ LL remains an open question and will the subject of an upcoming publication.

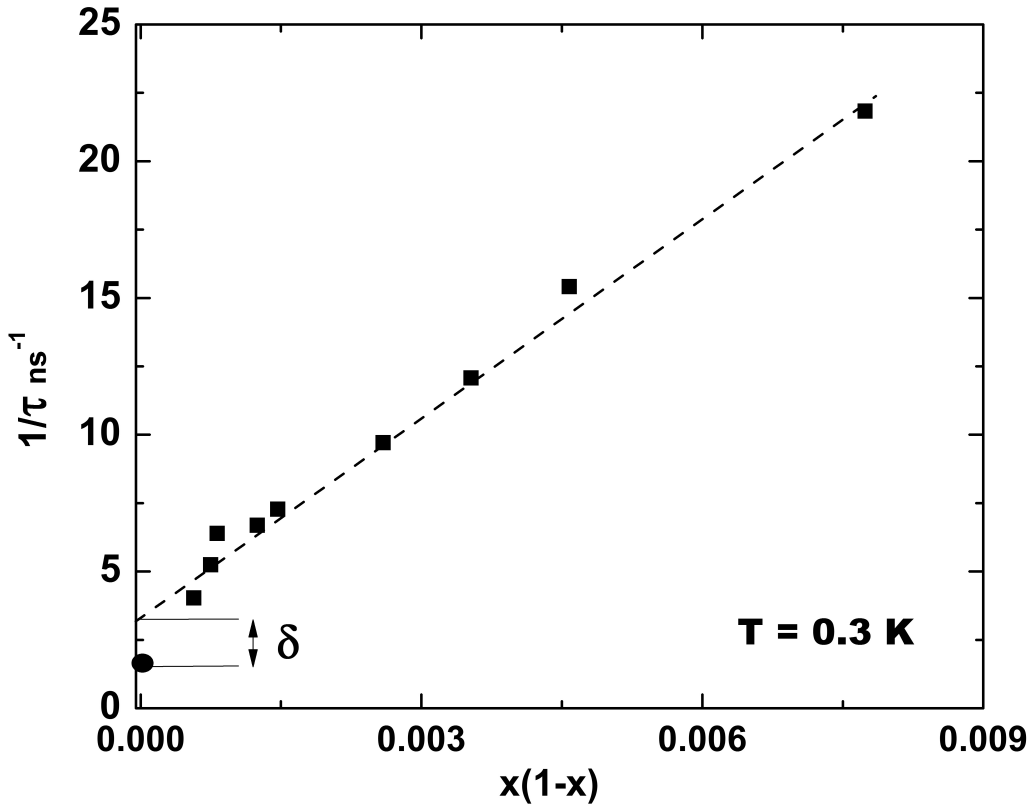

Figure 3: The dependence of $\tau^{-1}$ on x(1-x) measured at T=0.3K  The dotted line is a linear fit to the data which yields 24 ns$^{-1}$ per %Al.

To conclude, we have demonstrated a repeatable and accurate method to grow high quality 2DEG's while varying, at extremely low concentrations, the amount of Al in an Al$_x$Ga$_{1-x}$As quantum well.  We generated 10 samples with varying alloy concentrations in the quantum well to study the dependence of the zero magnetic field scattering rate on alloy concentration.  We observe that the total scattering rate depends linearly on x(1-x) and the alloy scattering rate was determined to be 24 ns$^{-1}$ per % Al.  Alloy scattering becomes the dominant mobility limiting scattering mechanism for x>0.0005 for the heterostructure design employed here.  The samples and methods described here will be used in future studies of the impact of alloy disorder in the fractional quantum Hall regime.

**Acknowledgements**

M.J.M. acknowledges support from the Miller Family Foundation. The molecular beam epitaxy growth and transport measurements at Purdue are supported by the U.S. Department of Energy, Office of Basic Energy Sciences, Division of Materials Sciences and Engineering under Award DE-SC0006671.